\documentclass[multphys,vecphys]{svmult}

\usepackage{makeidx}     
\usepackage{graphicx}    
\usepackage{multicol}    

\usepackage{cite}

\makeindex             

\begin{document}

\title{Quantum Monte Carlo Simulations of
the Impurity-Induced Phase Transitions
in Low-Dimensional Magnets}
\titlerunning{Impurity-Induced Phase Transitions}

\author{Munehisa Matsumoto\inst{1}
\and Hajime Takayama\inst{2}}
\authorrunning{Munehisa Matsumoto and Hajime Takayama}

\institute{Department of Physics,
Graduate School of Science, Tohoku University,
Sendai 980-8578, Japan
\and
Institute for Solid State Physics,
University of Tokyo, Chiba 277-8581, Japan}
%
%

\maketitle


We study the impurity-induced phase transitions
in a quasi-one-dimensional Heisenberg
antiferromagnet doped with
magnetic spin-$1/2$ impurities
and non-magnetic ones. The impurity-induced
transition temperature determined by the quantum Monte Carlo method
with the continuous-time loop algorithm is
monotonically increasing as a function
of the magnitude of the impurity spin. To these results,
we give discussions based on the valence-bond solid-like picture for the
pure system and the inspection of the
local magnetic structure around the
impurities.

\section{Motivation}

Impurities in the low-dimensional
quantum magnets drastically
change the magnetic properties
of the host materials.
For the low-dimensional magnets
such as CuGeO$_3$~\cite{hase}
and PbNi$_2$V$_2$O$_8$~\cite{uchiyama},
pure systems have disordered ground states, while
non-magnetic-impurity-doped ground states are
antiferromagnetically ordered.
These impurity-induced ordered states
have been given a qualitative understanding
based on the valence-bond solid (VBS) picture~\cite{affleck}
for low-dimensional magnets. According to
the VBS picture,
a spin with the magnitude $S$ on each site is
treated as ferromagnetically coupled
$2S$ subspins with the magnitude $1/2$,
and the disordered ground state
of a low-dimensional antiferromagnet
can be seen as a state made of closely packed
spin-$1/2$ singlet pairs, which are called
valence bonds.
When a non-magnetic impurity is doped,
a valence bond is broken
after one of the spin in a pair is lost
and the counterpart subspin is liberated.
Thus non-magnetic impurities induce nearly-free local magnetic
moments around their doped sites
and there is long-range correlation between them
that keeps the bulk antiferromagnetic long-range order~\cite{yasuda}.

Recently, in the experiments using
the spin-$1$ gapped magnet PbNi$_2$V$_2$O$_8$,
the effects of several species of
magnetic impurities
were systematically investigated and it was found that
the impurity-induced transition temperatures show
non-monotonic dependence on the magnitude of the
impurity spin~\cite{imai}. In particular, compared with
the transition temperatures induced by
non-magnetic impurities,
those done by Cu$^{2+}$ are
found to be much lower.
These results can be thought to be reasonable if we
inspect the magnitude of the local magnetic moments
that are expected to appear
around impurities by the simplest picture
analogous to the VBS one. There are two spin-$1/2$
effective magnetic moments around one non-magnetic impurity,
and these are coupled by an effective ferromagnetic
coupling mediated by the interchain couplings,
making one spin-$1$ local magnetic moment.
For a magnetic spin-$1/2$ impurity, there is one
spin-$1/2$ local magnetic moment per one impurity.
Thus the magnitude of the local magnetic moments
is larger for the non-magnetic impurities than
for the magnetic spin-$1/2$ ones.
We can at least
hope to understand the experimental
result that the spin-$1/2$ impurities induce
much lower transition temparature than
the non-magnetic ones do.
Motivated by this observation, we do quantum Monte
Carlo simulations for the quasi-one-dimensional Heisenberg
antiferromagnets with site impurities and see if
our picture for the impurity-induced ordered state
is valid or not.

\section{Model and Method}
\label{model_et_method}

We introduce our model following the observation
described in the previous section.
The pure system is weakly coupled spin-$1$ antiferromagnetic
Heisenberg chains aligned
in parallel on a simple cubic lattice.
The Hamiltonian is written as follows.
\begin{equation}
{\cal H}=J\sum_{x,y,z}{\mathbf S}_{x,y,z}\cdot{\mathbf S}_{x+1,y,z}
+J'\left(\sum_{x,y,z}{\mathbf S}_{x,y,z}\cdot{\mathbf S}_{x,y+1,z}
+\sum_{x,y,z}{\mathbf S}_{x,y,z}\cdot{\mathbf S}_{x,y,z+1}\right)
\end{equation}
Here the spin operator ${\bf S}_{x,y,z}$ has the
magnitude $S\equiv|{\bf S}|=1$
and $x,y,z$ denote the points on a simple cubic lattice.
We consider only the nearest neighbor antiferromagnetic couplings.
The intrachain (interchain) coupling is denoted by $J(J')>0$.
We set the $x$ axis parallel to the chains.
The strength of the interchain coupling, $J'$,
is set to be $0.01J$. This value
is small enoungh to
allow the Haldane gap~\cite{haldane} to be finite even
in a three-dimensional space.
The critical point between the gapped phase
and the antiferromagnetically ordered one, $J'_{\rm c}$,
is estimated to be $J'_{\rm c}=0.013J$~\cite{sakai,yasuda2}.

In the impurity-doped system, the $S=1$ host spins
are randomly replaced with the $S\ne 1$ impurity spins
with the couplings between the spins left unchanged.
The aim of using this simple model is to
let the magnitude of local magnetic moments
play a dominant role in determining the impurity-induced
transition temperatures by letting all of the species
of impurities share the equal strength of the effective
couplings between the impurity neighborhoods
when the concentration of the impurities
is fixed. By this model we try to simulate the
impurity-induced transition in which
the non-magnetic impurities
yield higher transition temperature than the
magnetic spin-$1/2$ ones do.

The impurity-induced
transition temperatures are found as follows.
The physical observables are calculated by the quantum Monte Carlo
with the continuous-time loop algorithm~\cite{evertz,todo}.
The correlation length is calculated using the second moment 
of the two-point correlation function~\cite{todo,cooper}.
The ratio of the correlation length to the linear size
of the system is plotted with respect to the temperature
for several system sizes and the crossing point (if any is found)
gives the position of critical point. By this simple analysis,
enough accuracy in the data for the qualitative discussion in the present
study is obtained.
The concentration of impurities
is fixed to be $10\%$ and the impurity-induced transition
temperatures (if any is found) are compared with each other
between the species of impurities.

\section{Results and Discussions}
\label{results_et_discussions}

By the procedure described in
the previous section, we find
that all species of impurities induce
phase transitions in our model.
Specifically,
it is found that
non-magnetic impurities induce the N\'{e}el temperature
at $T_{\rm N}=0.03\pm 0.01$ and magnetic spin-$1/2$
ones do that at $T_{\rm N}=0.05\pm 0.01$. As for 
the other magnetic-impurity-induced transition temperatures,
we find that they are monotonically increasing as
a function of the magnitude
of the spin of impurities~\cite{mm}.
We must note that the
transition temperature induced
by the magnetic spin-$1/2$ impurities
is higher than that done by
the non-magnetic ones, which is in contrast with our
expectation before simulations.

\begin{figure}
\centering
\parbox[t]{.05\textwidth}{\vspace{-.3\textwidth}(a)}
\parbox[t]{.42\textwidth}{\includegraphics[width=.4\textwidth]{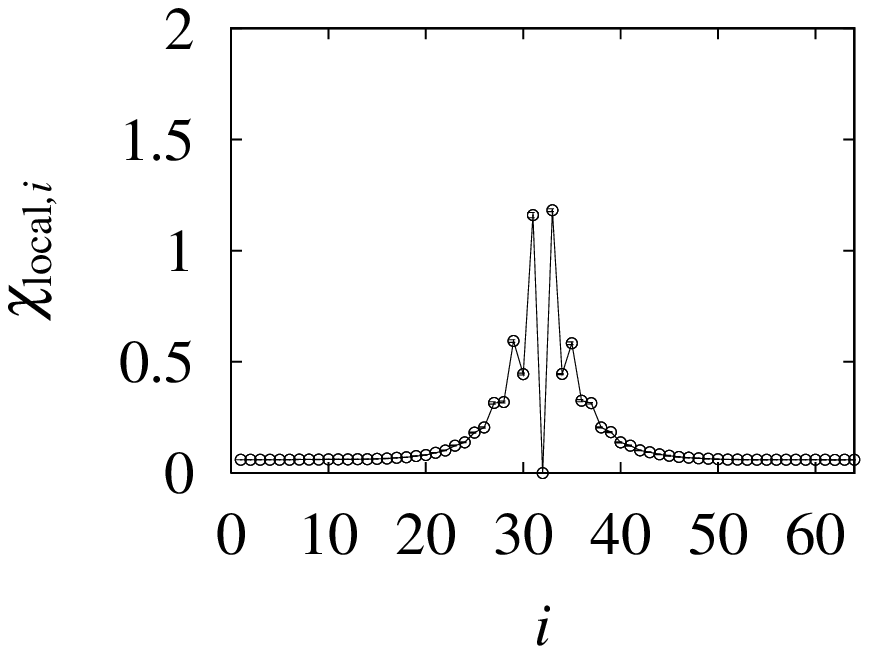}}
\parbox[t]{.05\textwidth}{\vspace{-.3\textwidth}(b)}
\parbox[t]{.42\textwidth}{\includegraphics[width=.4\textwidth]{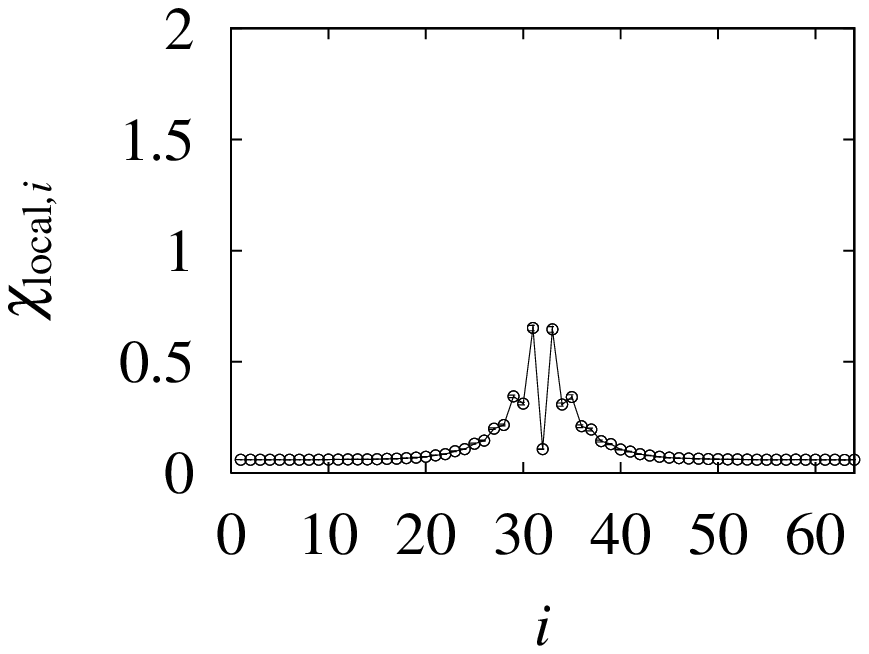}}
\caption{Local field susceptibilities plotted with respect to the site indices
of a spin-$1$ Haldane chain with an impurity doped in the center of it.
The length of the chain is $64$ and the temperature is $0.01$ measured by
the strength of the superexchange coupling.
An impurity is (a) non-magnetic and (b) magnetic with spin magnitude
$1/2$.}
\label{local}
\end{figure}

To check our presumption
that the magnitude of the local magnetic moments,
which should be relevant in the value of the
impurity-induced transition temperature,
must be larger for non-magnetic impurities than
for magnetic spin-$1/2$ ones,
we are going to take a look directly
on the local magnetic moments around an impurity.
We calculate the local field susceptilities
for each site $i$, which we denote by
$\chi_{{\rm local},i}$, on a single spin-$1$ chain
with one impurity doped in the center site
of the chain.
The local field susceptibility
is defined by the following formula.
\begin{equation}
\chi_{{\rm local},i}=
\left.\frac{\partial\left<S_{i}^z\right>}{\partial h_{i}}\right|_{h_i=0}=
\int_{0}^{\beta}d\,\tau
\left<S_{i}^{z}(\tau)S_{i}^{z}(0)\right>
\end{equation}
Here $h_i$ is the local magnetic field applied on the site $i$,
$\tau$ is the imaginary time introduced in the path-integral formalism
that is employed in the loop algorithm, and $\beta$ is the inverse temperature.
The real-space distribution of the
local field susceptibility is shown in Fig.~\ref{local}.
Here the length of the chain is $64$
and the temperature is $0.01J$ at which
the pure system is well near the ground state.
We see that the total magnitude of
the local magnetic moments is larger
in the neighborhood of a non-magnetic impurity
than in that of a magnetic spin-$1/2$ one.
This result is
consistent with our first expectation.

\begin{figure}
\centering
\parbox[t]{.05\textwidth}{\vspace{-.3\textwidth}(a)}
\parbox[t]{.42\textwidth}{\includegraphics[width=.4\textwidth]{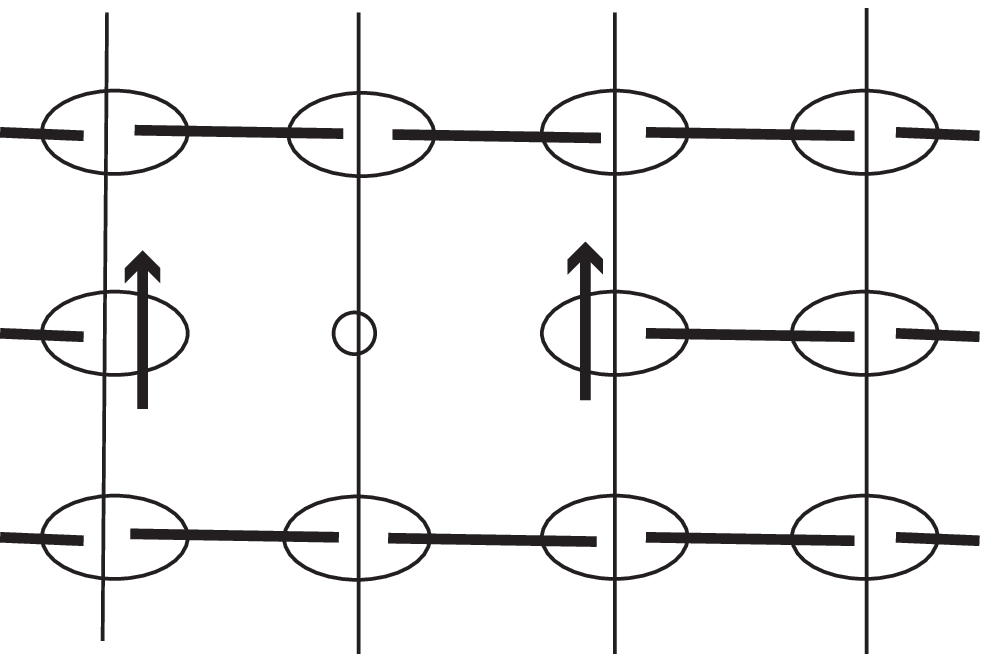}}
\parbox[t]{.05\textwidth}{\vspace{-.3\textwidth}(b)}
\parbox[t]{.42\textwidth}{\includegraphics[width=.4\textwidth]{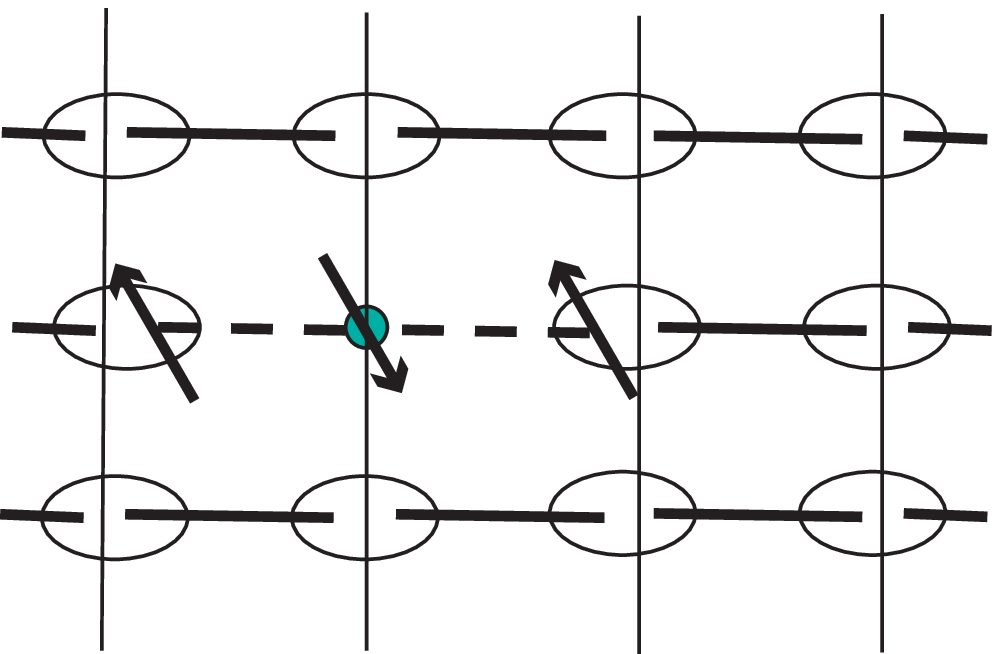}}
\caption{Schematic figure for the local spin-spin couplings between
local magnetic moments around (a) a non-magnetic impurity
and (b) a spin-$1/2$ impurity in the gapped phase of the quasi-one-dimensional
spin-$1$ Heisenberg antiferromagnet. The bold solid line represents a valence bond
and the dotted line do the intrachain coupling.
The arrows denote local spin-$1/2$ effective
magnetic moments.}
\label{local_schematic}
\end{figure}

As the transition temperature is
expected to be determined by both of the magnitude of
the magnetic moments that contribute to the long-range order
and the strength of the interaction between them,
we are led to consider the interaction
between the local magnetic moments more closely.
First we consider the neighborhoods of non-magnetic impurities.
The two local magnetic moments with the effective magnitude
$1/2$ around a non-magnetic impurity are ferromagnetically
coupled mediated by the very weak interchain couplings.
The strength of this coupling
is estimated to be of the order $J'^2$.
Near the transition temperature that is in energy scale
comparable to the strength of the interchain coupling $J'$,
the coupling of the order $J'^2$ between the two moments hardly
contribute to the global antiferromagnetic order of the present
interest.
On the other hand, around
a magnetic spin-$1/2$ impurity, two effective spin-$1/2$ moments
at the edge of the pure chain and the impurity spin
are directly coupled
yielding an effective spin of the magnitude $1/2$. See
Fig.~\ref{local_schematic}. When the ground state
of the three spins with the magnitude $1/2$ which are
separated from the rest is inspected, they exhibit magnetic moments
(whose magnitudes are less than $1/2$) antiferromagnetically aligned.
This strongly suggests that the three spins contribute to the global
antiferromagnetic order even at temperatures of the order $J'$, i.e.,
near the transition temperature.
Thus it is reasonable to encounter the present simulational
results that the impurity-induced temperatures are higher
for the magnetic impurities than for the non-magnetic ones.
Our first expectation may hold good in the ordering
at temperatures below the order $J'^2$.

In real experiments, various kinds of interaction
around impurities must be playing key roles,
which may require the detailed examination
specific to each kind of impurities~\cite{imai}.

\section*{Acknowledgements}
One of the authors (MM) thanks
Prof. T. Masuda and Prof. K. Uchinokura
for their useful comments and
Dr. C. Yasuda and Dr. S. Todo for
helpful discussions.
The computations for this work were
done on the SGI ORIGIN 2800 that was
at the Supercomputer center in the
Institute for Solid State Physics,
University of Tokyo until this February.
The codes are based
on the library ``LOOPER version 2'' developed by
Dr. S. Todo and Dr. K. Kato,
and ``PARAPACK version 2''
developed by Dr. S. Todo.


\begin{thebibliography}{99.}

\bibitem{hase} 
M. Hase, I. Terasaki, and K. Uchinokura:
Phys. Rev. Lett. \textbf{70}, 3651 (1993);
M. Hase, I. Terasaki, Y. Sasago,
K. Uchinokura, and H. Obara:
Phys. Rev. Lett. \textbf{71}, 4059 (1993).

\bibitem{uchiyama} Y. Uchiyama, Y. Sasago,
I. Tsukada, K. Uchinokura, A. Zheludev, T. Hayashi,
and, N. Miura, and P. B\"{o}ni:
Phys. Rev. Lett. \textbf{83}, 632 (1999).

\bibitem{affleck} I. Affleck, T. Kennedy, E.~H. Lieb,
and H. Tasaki: Phys. Rev. Lett. \textbf{59}, 799 (1987).

\bibitem{yasuda} C. Yasuda, S. Todo, M. Matsumoto, and H. Takayama:
Phys. Rev. B \textbf{64}, 092405 (2001).

\bibitem{imai} S. Imai, T. Masuda, T. Matsuoka, and K. Uchinokura:
preprint (cond-mat/0402595).

\bibitem{haldane}
F.~D.~M. Haldane, Phys. Lett. \textbf{A93}, 464 (1983);
Phys. Rev. Lett. \textbf{50}, 1153 (1983).

\bibitem{yasuda2} Recently the N\'{e}el temperatures
of quasi-low-dimensional antiferromagnets were investigated
systematically and modified random phase approximation theory
was discussed, in which the position of the
critical point between the gapped phase
and the long-range ordered phase can be estimated. 
See the following paper. C. Yasuda, S. Todo,
K. Hukushima, F. Alet, M. Keller, M. Troyer, and H. Takayama:
preprint (cond-mat/0312392).

\bibitem{sakai}
T. Sakai and M. Takahashi:
J. Phys. Soc. Jpn. \textbf{58}, 3131 (1989).

\bibitem{evertz}
H.~G. Evertz, G. Lana, and M. Marcu: Phys. Rev. Lett. \textbf{70}, 875 (1993).
B.~B. Beard and U.-J. Wiese: Phys. Rev. Lett. \textbf{77}, 5130 (1996).
Excellent reviews on the loop algorithm are given in the following
articles. H.~G. Evertz: Adv. Phys. \textbf{52}, 1 (2003);
N. Kawashima and K. Harada: J. Phys. Soc. Jpn. \textbf{73}, 1379 (2004).
See also the following article. K. Harada: this volume.

\bibitem{todo}
S. Todo and K. Kato: Phys. Rev. Lett. \textbf{87}, 047203 (2001).

\bibitem{cooper} F. Cooper, B. Freedman, and D. Preston:
Nucl. Phys. \textbf{B210[FS6]}, 210 (1982).

\bibitem{mm} M. Matsumoto and H. Takayama: preprint (cond-mat/0412676).



\end{thebibliography}
%



\printindex
\end{document}